\documentclass[final,numberedheadings,cmfonts]{aipproc}
\layoutstyle{6x9}
\usepackage{multirow}
\newcommand{\fr}{\frac}
\def\1{\mbox{l\hspace{-0.53em}1}}

\begin{document}

\title{Highly Excited Baryons in Large $N_c$ QCD}


\author{N. Matagne and Fl. Stancu}{address={University of Li\`ege, Physics Department,\\
Institute of Physics, B.5, \\ 
Sart Tilman, B-4000 Li\`ege 1, Belgium\\ 
E-mail: nmatagne@ulg.ac.be,
fstancu@ulg.ac.be}}

\begin{abstract}
We use the  $1/N_c$ expansion of QCD
to analyse the spectrum of positive parity resonances with strangeness
$S = 0, -1, -2$ and $-3$ in the 2--3 GeV mass region, supposed to belong 
to the $[\textbf{56},4^+]$ multiplet. The mass operator
is similar to that of  $[\textbf{56},2^+]$,
previously studied in the literature. The analysis of the latter is revisited. 
In the $[\textbf{56},4^+]$ multiplet we find that the spin-spin term brings the 
dominant contribution and that  the spin-orbit term is entirely negligible 
in the hyperfine interaction, in agreement with the constituent quark model practice, where this interaction is usually neglected. 
More data are strongly desirable, especially in the strange sector in order to 
fully exploit the power of this approach. We discuss possibilities of extending the calculations to other excited baryons belonging to the $N=2$ or the $N=4$ band.
\end{abstract}

\maketitle

\section{Introduction}

The $1/N_c$ expansion of QCD suggested by 't Hooft \cite{tHo74} about 30 years 
ago and analysed in a greater detail by Witten \cite{Wit79} has lead to a 
powerful algebraic method to study baryon spectroscopy. This method is based on 
the result that the SU(6) spin-flavor symmetry is exact in the large $N_c$ 
limit \cite{DM93}. It has been applied with a great success to the ground state 
baryons which correspond to the SU(6) {\bf 56} multiplet \cite{DJM94,DJM95,CGO94,Jenk1,JL95,DDJM96} as well as to some 
excited states, as for example the $[{\bf 70},1^-]$ negative parity states \cite{Goi97,PY,CCGL,CGKM,CaCa98,SGS,Pirjol:2003ye,CC00,GSS03,cohen1}.

The applicability of the large $N_c$ QCD to the description of excited baryon 
states is a problem of active investigation. Here we analyse the applicability 
of the method to the $[{\bf 56},4^+]$ multiplet and develop considerations for the treatment of multiplet as $[{\bf 70}, 0^+]$ or $[{\bf 70}, 2^+]$.

\section{The Mass Operator}

The study of the $[\textbf{56},4^+]$ multiplet is similar to that 
of $[\textbf{56},2^+]$
as analysed in Ref. \cite{GSS03}, where the mass
spectrum is calculated in the $1/N_c$ expansion up to and including 
$\mathcal{O}(1/N_c)$ effects.
 The SU(2) isospin symmetry is supposed to be exact.  
The SU(3) symmetry breaking is implemented 
to $\mathcal{O}(\varepsilon)$,
where $\varepsilon \sim 0.3$ gives a measure of this breaking by  the strange quark mass.
As the $\bf 56$  is a
  symmetric representation of SU(6), 
it is not necessary to distinguish between
excited and core quarks for the construction of a basis of mass operators,  
as explained in Ref. \cite{GSS03}. Then 
the mass operator of the SU(3) multiplets
has the following structure 
\begin{equation}
\label{MASS}
M = \sum_i c_i O_i + \sum_i b_i \bar{B}_i
\end{equation}
given in terms of the linearly independent operators $O_i$ and $\bar{B}_i$ defined in Table \ref{operators}. Here
$O_i$ ($i = 1,2,3$) are rotational invariants and SU(3) flavor singlets
\cite{Goi97}, $\bar{B}_1$ is the strangeness quark number operator
with negative sign, 
and the operators $\bar{B}_i$ ($i = 2, 3$)
are also rotational invariants but contain the SU(6) spin-flavor generators 
$G_{i8}$ as well. The operators
$\bar{B}_i$ ($i = 1, 2, 3$) provide SU(3) breaking and are defined to
have vanishing matrix elements for nonstrange baryons.
The relation (\ref{MASS}) contains the effective 
coefficients $c_i$ and $b_i$ as parameters. They represent reduced matrix 
elements that encode the QCD dynamics.
The values of the corresponding coefficients 
which we obtained from fitting the experimentally known masses  
are given in Table \ref{operators} both for the $[{\bf 56},4^+]$ and the 
presently revisited $[{\bf 56},2^+]$.

\section{The $[56,2^+]$ multiplet revisited}

As mentioned above, the study of the $[\textbf{56},4^+]$ multiplet is similar 
to that of $[\textbf{56},2^+]$. Here we first revisit 
the $[\textbf{56},2^+]$ multiplet for two purposes: 1) to get a consistency
test of our procedure of calculating matrix elements of the operators
in Table \ref{operators} and 2) to analyse a new assignement of
the $\Delta_{5/2^+}$ resonances. 

The matrix elements of $O_1$, $O_2$, $O_3$ and $\bar{B}_1$
are trivial to calculate for both multiplets under study. 
For the $[{\bf  56},2^+]$ one can find them in Table 2 of Ref. \cite{GSS03}
and for the $[{\bf 56},4^+]$ they are given in the next section.

To calculate the diagonal and off-diagonal  matrix elements 
of $\bar{B}_2$ we use the definition
\begin{equation}\label{Gi8}
G_{i8} = G^{i8} = \frac{1}{2 \sqrt{3}}\left(S^i - 3 S^i_s\right), 
\end{equation}
where $S^i$ and $S^i_s$ are  components of the total spin 
and of the total strange-quark spin respectively \cite{JL95}. 
Using (\ref{Gi8}) we can rewrite ${\bar B}_2$ from Table  \ref{operators} as  
\begin{equation}\label{B2}
\bar{B}_2 = - \frac{\sqrt{3}}{2 N_c} \vec{l}\cdot \vec{S}_s
\end{equation}
with the decomposition 
\begin{equation}\label{decomposition}
\vec{l}\cdot \vec{S}_s = l_0S_{s0}+\frac{1}{2}\left( l_+S_{s-}+l_-S_{s+}\right).
\end{equation} 
We calculated the matrix elements from the wave functions   
used in constituent quark model studies, where the center of mass coordinate
has been removed and only the internal Jacobi coordinates appear
(see, for example, Ref. \cite{book}). The expressions we found
for the matrix elements of $\bar{B}_2$ were identical with those
of Ref. \cite{GSS03}, based on  Hartree wave functions,
exact in the $N_c \rightarrow \infty$ limit only. 
This proves that in the Hartree  approach no center of mass corrections are 
necessary  for the $[{\bf 56},2^+]$ multiplet. We expect the same conclusion 
to stand for any $[{\bf 56},\ell^+]$. For mixed representations
the situation is more intricate \cite{CCGL}, see section \ref{perspectives}.

For $\bar{B}_3$, we used the following 
relation \cite{CaCa98}
\begin{equation}\label{B3}
S_i G_{i8} = \fr{1}{4 \sqrt{2}} \left[3I(I+1) -S(S+1) - \fr{3}{4} N_s(N_s+2)\right]
\end{equation}
in agreement with \cite{JL95}. Here $I$ is the isospin, $S$ is the total
spin and  $N_s$ the number of strange quarks. As for the matrix elements of 
$\bar{B}_2$, we found identical results to those of Ref. \cite{GSS03}.  
Note that only  $\bar{B}_2$ has non-vanishing off-diagonal
matrix elements. Their role is very important in the state mixing in particular in the octet-decuplet mixing. We found that the diagonal matrix elements of $O_2$, $O_3$,
$\bar{B}_2$ and $\bar{B}_3$ of strange baryons satisfy the following relation
\begin{equation}\label{dependence}
\frac{\bar{B}_2}{\bar{B}_3} =\frac{O_2}{O_3},
\end{equation}
for any state, irrespective of the value of $J$ in both the octet and the
decuplet.  Such a relation also holds for the multiplet $[\textbf{56},4^+]$
studied in the next section and might possibly be a feature of all
$[\textbf{56},\ell^+]$ multiplets. It can be used as a check of the analytic expressions in Table \ref{octets}. In spite of the relation (\ref{dependence}) 
which holds for the diagonal matrix elements, the operators $O_i$ and $\bar{B}_i$ 
are linearly independent, as it can be easily proved. 

The other issue for the $[{\bf 56},2^+]$ multiplet is that the 
analysis performed in Ref. \cite{GSS03} is based on the standard 
identification of resonances due to the pioneering work of Isgur and Karl
\cite{IK}. In that work the spectrum of positive parity resonances was
calculated from a Hamiltonian containing a harmonic oscillator confinement
and a hyperfine interaction of one-gluon exchange type. The mixing angles
in the $\Delta_{5/2^+}$ sector turned out to be
\begin{center}
\renewcommand{\arraystretch}{1.25}
\begin{tabular}{ccc}
\hline \hline
State & Mass (MeV) &  Mixing angles\\
\hline
$^4\Delta[\textbf{56},2^+]\frac{5}{2}^+$ & $1940$ & \multirow{2}{2.95cm}{$\left[ \begin{array}{cc} 0.94  & 0.38 \\  -0.38 & 0.94 \end{array}\right]$} \\
$^4\Delta[\textbf{70},2^+]\frac{5}{2}^+$ & $1975$ & \\
\hline \hline
\end{tabular}
\end{center}
which shows that the lowest resonance at 1940 MeV is dominantly  
a $[\textbf{56},2^+]$ state. As a consequence, the lowest observed 
$F_{35}\ \Delta(1905)$ resonance was interpreted as a member of the $[\textbf{56},2^+]$ multiplet.

In a more realistic description, based on a linear confinement
\cite{SS86}, the structure of  the $\Delta_{5/2^+}$ sector appeared
to be different. The result was 

\begin{center}
\renewcommand{\arraystretch}{1.25}
\begin{tabular}{ccc}
\hline \hline
State & Mass (MeV) & Mixing angles \\
\hline
$^4\Delta[\textbf{56},2^+]\frac{5}{2}^+$ & 1962 & \multirow{2}{3.4cm}{$\left[ \begin{array}{cc} 0.408 & 0.913 \\ 0.913 & -0.408 \end{array}\right]$} \\
$^4\Delta[\textbf{70},2^+]\frac{5}{2}^+$ & $1985$ & \\
\hline \hline
\end{tabular}
\end{center}
which means that in this case the higher resonance, of mass 1985 MeV, is
dominantly $[\textbf{56},2^+]$. Accordingly, here we interpret the higher experimentally
observed resonance $F_{35}~ \Delta(2000)$ as belonging to the $[\textbf{56},2^+]$ multiplet
instead of the lower one. Thus we take as experimental input the mass
1976 $\pm$ 237 MeV, determined from the full listings of the PDG \cite{PDG04} 
in the same manner as for the one-
and two-star resonances of the $[\textbf{56},4^+]$
multiplet (see below). 
The fit for $[{\bf 56},2^+]$ multiplet based on this assignement is shown
in Table \ref{operators}.
The $\chi^2_{\mathrm{dof}}$ obtained is 0.58, as
compared to  $\chi^2_{\mathrm{dof}}$ = 0.7 of Ref. \cite{GSS03}. The contribution of the spin-orbit operator $O_2$ is slightly smaller here than  in Ref. \cite{GSS03}. 
Although $\Delta(2000)$ is a two-star resonance only, the incentive of
making the above choice was that the calculated pion decay 
widths of the $\Delta_{5/2^+}$ sector were better reproduced \cite{SS95} 
with the mixing angles of the
model \cite{SS86} than with those of the standard model of Ref. \cite{IK}.  
It is well known that decay widths are useful to test mixing angles. 
Moreover, it would be more natural that the resonances $\Delta_{1/2}$
and $\Delta_{5/2}$ would have different masses, contrary to the 
assumption of Ref. \cite{GSS03} where these masses were taken to be identical.

\begin{table}[ph]
\caption{Operators of Eq. (\ref{MASS}) and coefficients resulting from the fit with
$\chi^2_{\rm dof}  \simeq 0.58$ for $[{\bf 56},2^+]$ and
$\chi^2_{\rm dof}  \simeq 0.26$ for $[{\bf 56},4^+]$.}
{\footnotesize
\renewcommand{\arraystretch}{1.25}
\begin{tabular}{@{}llrll@{}}
\hline
\hline
{} &{} &{} &{} &{}\\[-1.5ex]
{} Operator & \multicolumn{4}{c}{Fitted coef. (MeV)} \\[1ex]
\hline
{} &{}  $[{\bf 56},2^+]$  &{} &{} $[{\bf 56},4^+]$ &{}\\[1ex]
\hline
\hline
{} &{} &{} &{} &{}\\[-1.5ex]
$O_1 = N_c \ \1 $  &{}  $c_1$ = 540 $\pm$ 3 &{} &{}  $c_1$ = 736  $\pm$  30 &{}\\[-1.5ex]
{} &{} &{} &{} &{}\\[-1.5ex]
$O_2 =\frac{1}{N_c} l_i S_i$ &{} $c_2$ = 14 $\pm$ 9 &{} &{} $c_2 =$ 4 $\pm$ 40 &{}\\[-1.5ex]
{} &{} &{} &{} &{}\\[-1.5ex]
$O_3 = \frac{1}{N_c}S_i S_i$ &{} $c_3$ = 247 $\pm$ 10 &{} &{} $c_3$ = 135 $\pm$ 90 &{}\\[-1.5ex]
{} &{} &{} &{} &{}\\[-1.5ex]
{} &{} &{} &{} &{}\\[-1.5ex]
\hline
{} &{} &{} &{} &{}\\[-1.5ex]
$\bar{B}_1 = - \mathcal{S} $ &{} $b_1$ = 213 $\pm$ 15 &{} &{} $b_1$ = 110 $\pm$ 67 &{}\\[-1.5ex]
{} &{} &{} &{} &{}\\[-1.5ex]
$\bar{B}_2 = \frac{1}{N_c} l_i G_{i8}-\frac{1}{2 \sqrt{3}} O_2$  &{} 
$b_2$ = 83 $\pm$ 40 &{} &{} &{}\\[-1.5ex]
{} &{} &{} &{} &{}\\[-1.5ex]
$\bar{B}_3 = \frac{1}{N_c} S_i G_{i8}-\frac{1}{2 \sqrt{3}} O_3$  &{}  
$b_3$ = 266 $\pm$ 65 &{} &{} &{}\\[-1.5ex]
{} &{} &{} &{} &{}\\[-1.5ex]
{} &{} &{} &{} &{}\\[-1.5ex]
\hline \hline
\end{tabular}\label{operators} }
\vspace*{-13pt}
\end{table}

\section{The $[56,4^+]$ multiplet}

Tables \ref{singlets} and \ref{octets}  give all matrix elements needed for the octets and
decuplets belonging to the $[{\bf 56},4^+]$  multiplet. They are calculated 
following the prescription of the previous section. 
This means that the matrix elements of $O_1$, $O_2$, $O_3$ and $\bar{B}_1$
are straightforward and for $\bar B_3$ we use the formula (\ref{B3}).
The  matrix elements of $\bar B_2$ are calculated from the 
wave functions given explicitly
in Ref. \cite{MS}, firstly derived and employed in constituent quark model
calculations \cite{SS95}. One can see that the relation (\ref{dependence})
holds for this multiplet as well.

As mentioned above, only the operator $\bar{B}_2$ has non-vanishing off-diagonal 
matrix elements, so $\bar{B}_2$ is the only one which 
induces mixing between the octet and decuplet states of $[\textbf{56},4^+]$
with the same quantum numbers, as a consequence of the SU(3) flavor breaking.
Thus this mixing affects the octet and the decuplet $\Sigma$ and 
$\Xi$ states. As there are four off-diagonal matrix elements
(Table \ref{octets}), there are also 
four mixing angles, namely,  $\theta_J^{\Sigma}$ and $\theta_J^{\Xi}$, each with
$J = 7/2 $ and 9/2.  In terms of these mixing angles, the physical $
\Sigma_J$ and $\Sigma_J'$ states are defined by the following basis states
\begin{eqnarray}
|\Sigma_J \rangle  & = & |\Sigma_J^{(8)}\rangle \cos\theta_J^{\Sigma} 
+ |\Sigma_J^{(10)}\rangle \sin\theta_J^{\Sigma}, \\
|\Sigma_J'\rangle & = & -|\Sigma_J^{(8)}\rangle \sin\theta_J^{\Sigma}
+|\Sigma_J^{(10)} \rangle \cos\theta_J^{\Sigma},
\end{eqnarray}
and similar relations hold for $\Xi$. 
The masses of the physical states become
\begin{eqnarray}
M(\Sigma_J) & = & M(\Sigma_J^{(8)}) 
+ b_2 \langle \Sigma_J^{(8)}|\bar{B}_2|\Sigma_J^{(10)} \rangle \tan \theta^{\Sigma}_J, \label{masssigmaj} \\
M(\Sigma_J') & = & M(\Sigma_J^{(10)}) 
- b_2 \langle \Sigma_J^{(8)}|\bar{B}_2|\Sigma_J^{(10)} \rangle \tan \theta^{\Sigma}_J \label{masssigma'j},
\end{eqnarray}
where $M(\Sigma_J^{(8)})$ and $M(\Sigma_J^{(10)})$ are the diagonal matrix 
of the mass operator (\ref{MASS}), 
here equal to $c_1O_1+c_2O_2+c_3O_3+b_1\bar{B}_1+b_2\bar{B}_2+b_3\bar{B}_3$, 
for $\Sigma$ states and  similarly for $\Xi$ states 
(see Table \ref{multiplet}). 
If replaced in the mass operator (\ref{MASS}), the relations (\ref{masssigmaj}) 
and  (\ref{masssigma'j}) and their counterparts for $\Xi$, introduce 
four new parameters which should be included in the fit.
Actually the procedure of Ref. \cite{GSS03} was simplified to fit the coefficients 
$c_i$ and $b_i$ directly to the physical masses
and then to calculate the mixing angle from
\begin{equation}
\theta_J = 
\frac{1}{2}\arcsin\left( 2~ \frac{b_2\langle \Sigma_J^{(8)}|\bar{B}_2|\Sigma_J^{(10)} \rangle}
{ M(\Sigma_J) - M(\Sigma'_J)}\right).
\label{mixingangles}
\end{equation} 
for $\Sigma_J$ states and analogously for $\Xi$ states.  

Due to the scarcity of
data in the 2--3 GeV mass region, even such a simplified procedure
is not possible at present in the $[\textbf{56},4^+]$ multiplet.

The fit of the  masses derived from  Eq. (\ref{MASS})
and the available empirical values used in the fit,
together with the corresponding resonance status in the Particle Data Group
\cite{PDG04} are listed in Table \ref{multiplet}.
The values of the coefficients $c_i$ and $b_1$ obtained from 
the fit are presented in Table \ref{operators}, as already mentioned.
For the four and three-star resonances we used the empirical masses given in 
the summary  table. For the others, namely the one-star resonance
$\Delta(2390)$ and the two-star resonance $\Delta(2300)$
we adopted the following procedure. We considered as ``experimental'' mass
the average of all masses quoted in the full listings. The experimental
error to the mass was defined as the quadrature of two uncorrelated errors,
one being the average error obtained from the same references in the 
full listings  and the other was the difference 
between the average mass relative to the farthest off observed mass. The 
masses and errors thus obtained are indicated in the before last column
of Table \ref{multiplet}.

\begin{table}
\caption{Matrix elements of  SU(3) singlet operators.}
{\renewcommand{\arraystretch}{1.75}
\begin{tabular}{@{}crrr@{}}
\hline
\hline
           & $\ \ \ \ \ \ \ \  O_1$  & $\ \ \ \ \ \ \ O_2$  & $\ \ \ \ \ \ \ O_3$  \\
\hline
$^28_{7/2}$ & $N_c$  & $- \fr{5}{2 N_c}$ & $\fr{3}{4 N_c}$  \\
$^28_{9/2}$  & $N_c$  &   $\fr{2}{  N_c}$ & $\fr{3}{4 N_c}$  \\
$^410_{5/2}$ & $N_c$  & $- \fr{15}{2 N_c}$ & $\fr{15}{4 N_c}$ \\
$^410_{7/2}$ & $N_c$  & $- \fr{4}{  N_c}$ & $\fr{15}{4 N_c}$ \\
$^410_{9/2}$ & $N_c$  &   $\fr{1}{2 N_c}$ & $\fr{15}{4 N_c}$ \\
$^410_{11/2}$ & $N_c$  &   $\fr{6}{  N_c}$ & $\fr{15}{4 N_c}$ \\[0.5ex]
\hline
\hline
\end{tabular}}

\label{singlets}
\end{table}

\begin{table}
\caption{Matrix elements of SU(3) breaking operators. Here, $a_J = 5/2,-2$ for $J=7/2, 9/2$, respectively and
$b_J = 5/2, 4/3, -1/6, -2$ for $J=5/2, 7/2, 9/2, 11/2$, respectively.}
{\renewcommand{\arraystretch}{1.75}
\begin{tabular}{cccc}
\hline
\hline
  &  \hspace{ .6 cm}  ${\bar B}_1$ \hspace{ .6 cm}  &
     \hspace{ .6 cm}  ${\bar B}_2$ \hspace{ .6 cm}  &
     \hspace{ .6 cm}  ${\bar B}_3$ \hspace{ .6 cm}   \\
\hline
$N_{J}$       & 0 &                      0  &                           0  \\
$\Lambda_{J}$ & 1 &   $\fr{  \sqrt{3}\ a_J}{2 N_c}$ &   $- \fr{3 \sqrt{3}}{8 N_c}$   \\
$\Sigma_{J}$  & 1 & $- \fr{  \sqrt{3}\ a_J}{6 N_c}$ &     $\fr{  \sqrt{3}}{8 N_c}$   \\
$\Xi_{J}$     & 2 &   $\fr{ 2\sqrt{3}\ a_J}{3 N_c}$ &   $- \fr{  \sqrt{3}}{2 N_c}$   \\ [0.5ex]
\hline
$\Delta_{J}$  & 0 &                      0  &                           0  \\
$\Sigma_{J}$  & 1 &  $\fr{ \sqrt{3}\ b_J}{2 N_c}$ &   $- \fr{ 5 \sqrt{3}}{8 N_c}$  \\
$\Xi_{J}$     & 2 &  $\fr{ \sqrt{3}\ b_J}{  N_c}$ &   $- \fr{ 5 \sqrt{3}}{4 N_c}$  \\
$\Omega_{J}$  & 3 &  $\fr{3\sqrt{3}\ b_J}{2 N_c}$ &   $- \fr{15 \sqrt{3}}{8 N_c}$  \\ [0.5ex]
\hline
$\Sigma_{7/2}^8$ $- \Sigma^{10}_{7/2}$    &  0 &  $-\fr{  \sqrt{35}}{2 \sqrt{3} N_c}$   &   0   \\
$\Sigma_{9/2}^8 - \Sigma^{10}_{9/2}$    &  0 &  $-\fr{  \sqrt{11}}{  \sqrt{3} N_c}$   &   0   \\
$\Xi_{7/2}^8    - \Xi^{10}_{7/2}$       &  0 &  $-\fr{  \sqrt{35}}{2 \sqrt{3} N_c}$   &   0   \\
$\Xi_{9/2}^8    - \Xi^{10}_{9/2}$       &  0 &  $-\fr{  \sqrt{11}}{  \sqrt{3} N_c}$   &   0   \\[0.5ex]
\hline
\hline
\end{tabular}}
\label{octets}
\end{table}

Due to the lack of experimental data in the strange sector it was 
not possible to include all the operators $\bar{B}_i$ in the fit in order 
to obtain 
some reliable predictions.  As the breaking of SU(3) is dominated by 
$\bar{B}_1$ we included only this operator in  Eq. (\ref{MASS})
and neglected the contribution of the operators $\bar{B}_2$ and $\bar{B}_3$.
At a later stage, when more data will hopefully be available, all analytical 
work performed here could be used to improve the fit. That is why
Table \ref{operators} contains results for  $c_i$ ($i$ = 1, 2 and 3) 
and $b_1$ only. The $\chi^2_{\mathrm{dof}}$ of the fit is 0.26, where 
the number of degrees of freedom (dof) is equal to one (five data and four 
coefficients).

\begin{table}
\caption{Masses (in MeV) of the $[{\bf 56},4^+]$ multiplet 
predicted by the $1/N_c$ expansion as compared with the 
empirically known masses. The partial contribution of each operator is indicated for
all masses. Those partial contributions in blank are equal to the one 
above in the same column.}
{\footnotesize
\renewcommand{\arraystretch}{1.5}
\begin{tabular}{crrrrccl}\hline \hline
                    &      \multicolumn{6}{c}{1/$N_c$ expansion results}        &                     \\ 
\cline{1-6}		    
                    &      \multicolumn{4}{c}{Partial contribution (MeV)} &  Total (MeV)   &   Empirical  &  Name, status \\
\cline{2-5}
                    &     $c_1O_1$  &   $c_2O_2$ & $c_3O_3$ &  $b_1\bar B_1$   &  &  (MeV)   &    \\
\hline
$N_{7/2}$        & 2209 & -3 &  34 &   0  &  $ 2240\pm97 $ &  &  \\
$\Lambda_{7/2}$  &      &     &    & 110  &  $2350\pm118 $  & & \\
$\Sigma_{7/2}$   &      &     &    & 110  &  $2350\pm118 $  &               & \\
$\Xi_{7/2}$      &      &     &    & 220  &  $2460\pm 166$  &               &  \\
\hline
$N_{9/2}  $      & 2209 & 2   & 34 &   0  &   $2245\pm95 $  & $ 2245\pm65 $ & N(2220)**** \\
$\Lambda_{9/2}$  &  &     &    & 110  &  $ 2355\pm116 $  & $ 2355\pm15 $ &  $\Lambda$(2350)***\\
$\Sigma_{9/2}$   &      &     &    & 110  &  $ 2355\pm116 $  &               &                    \\
$\Xi_{9/2}$      &      &     &    & 220  &  $2465\pm164$  &               &                    \\
\hline
$\Delta_{5/2}$   & 2209 & -9  &168 &   0  &  $ 2368\pm175$  & &  \\
$\Sigma^{}_{5/2}$&      &     &    & 110  & $2478\pm187$  & &   \\
$\Xi^{}_{5/2}$   &      &     &    & 220  &  $2588\pm220$  &               &                   \\
$\Omega_{5/2}$   &      &     &    & 330  & $2698\pm266$  &               &                    \\
\hline
$\Delta_{7/2}$   &2209  &-5   &168 &  0   &  $2372\pm153$  & $2387\pm88$ &  $\Delta$(2390)* \\
$\Sigma'_{7/2}$  &      &     &    & 110  & $2482\pm167$  &               &                  \\
$\Xi'_{7/2}$     &      &     &    & 220  & $2592\pm203$  &               &                    \\
$\Omega_{7/2}$   &      &     &    & 330  &  $2702\pm252$  &               &                    \\
\hline
$\Delta_{9/2}$   &2209  & 1   &168 &  0   &   $2378\pm144 $  & $2318\pm132  $ &  $\Delta$(2300)**\\
$\Sigma'_{9/2}$  &      &     &    & 110  &   $2488\pm159$  &               &                   \\
$\Xi'_{9/2}$     &      &     &    & 220  &  $2598\pm197$  &               &                    \\
$\Omega_{9/2}$   &      &     &    & 330  &  $2708\pm247$  &               &                    \\
\hline
$\Delta_{11/2}$  &2209  &7    &168 &  0   &  $2385\pm164$  & $ 2400\pm100$ &   $\Delta$(2420)**** \\
$\Sigma^{}_{11/2}$ &    &     &    & 110  & $2495\pm177$  &               &                     \\
$\Xi^{}_{11/2}$  &      &     &    & 220  &  $2605\pm212$  &               &                     \\
$\Omega_{11/2}$  &      &     &    & 330  &  $2715\pm260$  &               &                     \\
\hline
\hline
\end{tabular}}

\label{multiplet}
\end{table}

The first column of Table  \ref{multiplet}
 contains the 56 states (each state having a $2 I + 1$ multiplicity
from assuming an exact SU(2) isospin symmetry\footnote{%
Note that the notation $\Sigma_J$, $\Sigma'_J$ is consistent with 
the relations (\ref{masssigmaj}),~(\ref{masssigma'j}) inasmuch as 
the contribution of $\bar{B}_2$ is neglected (same remark 
for $\Xi_J$, $\Xi'_J$ and corresponding relations).}). 
The columns two to five show the 
partial contribution of each operator included in the fit, multiplied by
the corresponding coefficient $c_i$ or $b_1$.
The column six gives the total mass according to Eq. (\ref{MASS}). 
The errors shown in the predictions result from the errors on the 
coefficients $c_i$ and $b_1$ given in Table \ref{operators}.
As there are only five experimental data available, nineteen of these 
masses are predictions. The breaking of SU(3) flavor due to the operator
$\bar B_1$ is 110 MeV as compared to 200 MeV produced in the 
$[\textbf{56},2^+]$ multiplet.


Our results can be can be summarized as follows:
\begin{itemize}
\item The main part of the mass is provided by the spin-flavor singlet operator
$O_1$, which is $\mathcal{O}(N_c)$. 

\item The spin-orbit contribution given by $c_2O_2$ is small. This fact 
reinforces the practice used in constituent quark models where the spin-orbit
contribution is usually neglected in order to obtain a good fit. It is also consistent with the intuitive picture of Ref. \cite{glozman} where the spin-orbit interaction vanishes at high excitation energy. 
  
\item The breaking of the SU(6) symmetry keeping the flavor symmetry exact
is mainly due to the spin-spin operator $O_3$. This hyperfine interaction 
produces a splitting between octet and decuplet states of approximately 130 MeV 
which is smaller than that obtained in the $[\textbf{56},2^+]$ 
case \cite{GSS03}, which gives 240 MeV.

\item The contribution of $\bar{B}_1$
per unit of strangeness, 110 MeV, is also smaller here than in the 
$[\textbf{56},2^+]$ multiplet \cite{GSS03}, where it takes a value of about 
200 MeV. That may be quite natural, as one expects a shrinking of the spectrum
with the excitation energy.

\item As it was not possible to include the contribution of $\bar{B}_2$ and 
$\bar{B}_3$ in our fit, a degeneracy appears between $\Lambda$ and $\Sigma$. 

\end{itemize}


In conclusion we have studied the spectrum of highly excited resonances in the 2--3 GeV
mass region by describing them as belonging to the $[\textbf{56},4^+]$ multiplet.
This is the first study of such excited states based on the $1/N_c$ 
expansion of QCD. A better description should include multiplet 
mixing, following the lines developed, for example, in Ref. \cite{Go04}. 
 
We support previous assertions that better experimental values 
for highly excited non-strange baryons as well as more data 
for the $\Sigma^*$ and $\Xi^*$ baryons are needed in order to understand
the role of the operator $\bar{B}_2$  within a multiplet
and for the octet-decuplet mixing. With better data the analytic work 
performed here will help 
to make reliable predictions in the large $N_c$ limit formalism.

\section{Perspectives}
\label{perspectives}
As clearly stated in Ref. \cite{GSS03} the study of excited baryons is not free of difficulties. The use of the spin-flavor symmetry at zero-th order can be justified only from practical point of view, due to the smallness of the spin-orbit effects and of configuration mixings. More fundamentally, the excited baryons are unstable states for which the consistency condition is more difficult to ensure in large $N_c$ QCD \cite{cohen2}.

Moreover the analysis of the multiplet $[{\bf 56}, \ell^+]$  is simplified by the fact that a distinction between excited and core quarks is not necessary. This is not the case for mixed representations.

Our next objective is to analyse the $[{\bf 70},0^+]$ and $[{\bf 70},2^+]$ multiplets. The simplicity of the  $[{\bf 56}, \ell^+]$ does not hold anymore. In addition, one can not assume that the excitation is associated to a single quark which can be decoupled from a core free of excitations, as in the case of $[{\bf 70},1^-]$.

This can be illustrated by writing, for example, the two independent wave functions associated to $[{\bf  70},2^+]$ in the form where the center of mass motion has been removed. For $N_c=3$ we have
\begin{equation}
|{\bf 70}, 2^+\rangle_{\rho,\lambda} =  \sqrt{\frac{1}{3}}|[21]_{\rho,\lambda}(0s)^2(0d)\rangle +\sqrt{\frac{2}{3}}|[21]_{\rho,\lambda}(0s)(0p)^2\rangle.
\end{equation}
The coefficients of the linear combination above are independent of $N_c$ (we could prove this assertion for $N_c = 4, 5$ and 6). This implies that the two terms contribute to the same order and we have to consider both of them in the calculations.
The first is common to $[{\bf 56},2^+]$ and will be treated accordingly. For the second we have to include excitations both in the core and the decoupled quark. We shall analyse the role of an excited core in a future publication.

\section*{Acknowledgments}

The work of one of us (N. M.) was supported by the Institut Interuniversitaire 
des Sciences Nucl\'eaires (Belgium).


\begin{thebibliography}{0}

\bibitem{tHo74} G. 't Hooft,  Nucl. Phys. B {\bf 72} (1974) 461.

\bibitem{Wit79} E. Witten,  Nucl. Phys. B {\bf 160} (1979) 57.



\bibitem{DM93}
R.~Dashen and A.~V.~Manohar,
 Phys.~Lett. B {\bf 315} (1993) 425;
Phys.~Lett. B {\bf 315} (1993) 438.
\bibitem{DJM94}
R.~Dashen, E.~Jenkins, and A.~V.~Manohar,
Phys.~Rev.  D {\bf 49} (1994) 4713.
\bibitem{DJM95}
R. Dashen, E. Jenkins, and A. V. Manohar,
 Phys. Rev. D {\bf 51} (1995) 3697.
\bibitem{CGO94}
C.~D.~Carone, H.~Georgi and S.~Osofsky,
 Phys.~Lett. B {\bf 322} (1994) 227.\\
M.~A.~Luty and J.~March-Russell,
 Nucl.~Phys. B {\bf 426} (1994) 71.\\
M.~A.~Luty, J.~March-Russell and  M.~White,
 Phys.~Rev. D {\bf 51} (1995)  2332.
\bibitem{Jenk1}
E.~Jenkins,  Phys.~Lett. B  {\bf 315} (1993) 441.
\bibitem{JL95}
E.~Jenkins and R.~F.~Lebed,
 Phys.~Rev. D {\bf 52} (1995) 282.
\bibitem{DDJM96}
J.~Dai, R.~Dashen, E.~Jenkins, and A.~V.~Manohar,
 Phys.~Rev. D   {\bf 53} (1996) 273.
\bibitem{Goi97}
J.~L. Goity,
 Phys. Lett. B {\bf 414}, (1997) 140. 
\bibitem{PY}
D. Pirjol and T.~M. Yan,
 Phys. Rev. D {\bf 57} (1998) 1449. \\
D. Pirjol and T.~M. Yan,
 Phys. Rev. D {\bf 57} (1998) 5434.
\bibitem{CCGL}
C.~E. Carlson, C.~D. Carone, J.~L. Goity and R.~F. Lebed,
 Phys. Lett. B {\bf 438} (1998) 327;
 Phys. Rev. D {\bf 59} (1999) 114008.
\bibitem{CGKM}
C.~D.~Carone, H.~Georgi, L.~Kaplan and D.~Morin,
 Phys.\ Rev.\  D {\bf 50} (1994) 5793. 
\bibitem{CaCa98}
C.~E.~Carlson and C.~D.~Carone,
 Phys.\ Lett.\  B {\bf 441} (1998) 363;
 Phys.\ Rev.\  D {\bf 58} (1998) 053005.
\bibitem{SGS}
C.~L. Schat, J.~L. Goity and N.~N. Scoccola,  Phys. Rev. Lett.
{\bf 88} (2002) 102002;
J.~L.~Goity, C.~L.~Schat and N.~N.~Scoccola,
 Phys.\ Rev.\ D {\bf 66} (2002) 114014.
\bibitem{Pirjol:2003ye}
D.~Pirjol and C.~Schat,
 Phys. Rev. D {\bf 67} (2003) 096009.
\bibitem{CC00}
C. E. Carlson and C. D. Carone,
 Phys. Lett. B {\bf 484} (2000) 260.
\bibitem{GSS03}
J.~L. Goity,   C.~L.~Schat and N.~N.~Scoccola,
 Phys. Lett. B {\bf 564} (2003) 83.

\bibitem{cohen1}
T. D. Cohen {\it et al.}, Phys Rev D {\bf 69} (2004) 056001.

\bibitem{book} Fl. Stancu, {\it Group Theory in Subnuclear Physics}, Clarendon
Press, Oxford, 1996, Ch. 10.

\bibitem{IK} 
N. Isgur and G. Karl,  Phys. Rev.  D {\bf 19} (1979) 2653.

\bibitem{SS86} 
R. Sartor and Fl. Stancu, Phys. Rev. D {\bf 34} (1986) 3405.


\bibitem{PDG04}
S. Eidelman et al.,  Phys. Lett. B {\bf 592} (2004) 1.

\bibitem{SS95} 
P. Stassart and Fl. Stancu,   Z. Phys. A {\bf 351} (1995) 77.

\bibitem{MS}
N. Matagne and Fl. Stancu, Phys. Rev.  D {\bf 71} (2005) 014010. 

\bibitem{glozman}
  L.~Y.~Glozman,
  Phys.\ Lett.\ B {\bf 541} (2002) 115.

\bibitem{Go04} J.~L.~Goity,
  arXiv:hep-ph/0405304.

\bibitem{cohen2}
T. D. Cohen and R. F. Lebed, Phys. Rev. D {\bf 67} (2003) 096008; Phys. Rev. Lett {\bf 91} (2003) 012001; Phys. Rev. D {\bf 68} (2003) 056003. 




\end{thebibliography}
\end{document}